\documentclass[aps,pra,showpacs,groupedaddress,twocolumn]{revtex4}
\usepackage{amsfonts}
\usepackage{graphicx}
\usepackage{latexsym}
\usepackage{makeidx}
\usepackage{amsmath}
\usepackage{amssymb}

\begin{document}

\title{Approximate analytical results on the cavity dynamical Casimir effect
in the presence of a two-level atom}
\author{A. V. Dodonov and V. V. Dodonov}
\affiliation{Instituto de F\'{\i}sica, Universidade de Bras\'{\i}lia, PO Box 04455,
70910-900, Bras\'{\i}lia, DF, Brazil}

\begin{abstract}
We study analytically the photon generation from vacuum due to the Dynamical
Casimir effect in a cavity with a two-level atom, prepared initially in an
arbitrary pure state. Performing small unitary transformations we obtain
closed analytical expressions for the probability amplitudes and other
important quantities in the resonant/dispersive regimes.
\end{abstract}

\maketitle

\section{Introduction}

The fascinating dynamical Casimir effect (DCE), i.e., the creation of quanta
from vacuum due to the motion of macroscopic neutral boundaries (or due
to the time variations of material properties of these boundaries, such as the
dielectric constant or conductivity), attracted attention of many
theoreticians for several decades since the first publications
\cite{Moore,FulDav} (see \cite{revDCE,revDal,revDCE11} for the most recent reviews).
Quite recently the first experiments on modelling this effect in the superconducting
stripline waveguide terminated by a SQUID subjected to rapidly varying magnetic
flux (resulting in time-dependent boundary conditions simulating the motion of some
effective boundary) were performed \cite{DCE-Nature}. This realization can be
called the ``single mirror DCE'' \cite{FulDav,BE}. One of its specific features
(which was used as one of decisive proofs of the effect) is the creation of correlated
photon pairs emitted outside.

Another possible realization corresponds to the case
of a {\em closed\/} cavity with moving wall(s). This ``cavity DCE'' was considered
for the first time in \cite{Moore}, and it attracted the special attention,
because the
number of photons accumulated inside the cavity can be significantly increased in the case of {\em
periodical motion\/} of the wall(s) under certain resonance conditions
\cite{DK92,Law94,DK96}.
The simplest Hamiltonian describing this effect in the absense of dissipation reads
(we assume $\hbar=1$)
\cite{Law94}
\begin{equation}
H_{00}=\omega (t)n+i\chi (t)(a^{\dagger 2}-a^{2}),
\quad \chi(t) = \frac{d\omega /dt}{ 4\omega (t)},
\label{H00}
\end{equation}%
where $a$ and $a^{\dagger }$ are the cavity annihilation and creation operators,
$n\equiv a^{\dagger }a$ is the photon number operator, and $\omega (t)$ is the cavity
``instantaneous'' eigenfrequency, which depends on time due to the time-dependent
geometry of the cavity.
This Hamiltonian transforms the initial vacuum state into the {\em squeezed vacuum state},
so that only even numbers of quanta can be generated with nonzero probabilities.
Several possible realizations
of the cavity DCE Hamiltonian (\ref{H00})
were proposed a few years ago \cite{Padua05,Onof06}, and the experimental progress
was reported in \cite{Padova11} (other schemes, based on the fast optical modulation
of the cavity length, were proposed recently in \cite{optmod}).
In view of this progress, the problem of {\em detecting\/} the created
photons becomes a timely one.

One of the simplest detectors could be a two-level system (``atom'')
 \cite{PLA,Fedot00}, which can serve as an approximate model of either real Rydberg
atoms (or bunches of such atoms) \cite{Onof06,Kawa11} or some kinds of
``artificial atoms'' (made, e.g., from Josephson's contacts used
in quantum superconducting circuits \cite{artificial} -- in this case one deals with the
``circuit DCE'' \cite{revDCE11,circDCE}).
In such a case, one should add to Hamiltonian (\ref{H00})
the free atom Hamiltonian $H_{a}= (\Omega /2)\sigma_{z}$ and
the interaction Rabi Hamiltonian
\begin{equation}
H_{R}=
g\left(a+a^{\dagger }\right)\left(\sigma _{+}+\sigma _{-}\right),
\label{HR}
\end{equation}%
where $\sigma _{\pm }$ and $\sigma _{z}$ are the standard Pauli operators,
\[
\sigma _{z}=|e\rangle\langle e|-|g\rangle \langle g|, \quad \sigma _{-}=|g\rangle \langle e|,
\quad
\sigma _{+}=|e\rangle \langle g|.
\]
$\Omega $ and $g$ are the atomic transition frequency and the atom-field coupling
constant (assumed real), respectively.
The kets $|g\rangle $ and $|e\rangle $ can
be interpreted as \textquotedblleft atomic\textquotedblright\ ground and
excited states, respectively.

There are two possible scenarios. In the first one
the atoms are injected into the cavity after the walls made a sufficient number
of oscillations and returned to initial positions.
Then the solution is splitted in two steps: first one calculates how many
photons are created in the cavity, using Hamiltonian (\ref{H00}), and after
that one turns on the interaction between
the atoms and the field, using Hamiltonian
(\ref{HR}) together with the free field Hamiltonian with $\omega=const$.
This case was analyzed recently in \cite{Kawa11}.

Here we study another case: when the detecting atom is
present in the cavity for all time, influencing the photon generation
process. This situation looks quite realistic, because the cavities  in
the experiments proposed in \cite{Padua05,Kawa11} have the dimensions of the
order of a few centimeters (with the resonance frequencies of the order of a
few GHz, corresponding to the transitions between nearest levels of highly
excited Rydberg atoms). The travel time of atoms with velocities of the order
of $100\,$m/s through such cavities is of the order of $10^{-4}\,$s (or bigger
for slower atoms). On the other hand, the time of oscillations of the cavity
wall (more precisely, some effective ``moving plasma mirror'' created by periodic laser
pulses illuminating a thin semiconductor slab attached to the wall)
in the experiments discussed is expected to be of the order of $10^{-8}$--$10^{-6}\,$s,
so that it is much shorter than the atom travel time. Therefore it
could be more easy to send atoms through the cavity continuously, instead of
adjusting the exact moment of their entrance into the cavity. In such a case,
one can assume that the atom is permanently inside the cavity, thus interacting
with the field all time.
A similar situation can take place in the circuit DCE experiments, where
the artificial detecting ``atoms'' do not move at all (unless some scheme
of turning on their interaction with the field at precisely chosen instants of time is used).

Although this second scheme can result in diminishing the number of
created photons  \cite{PLA,Fedot00}, it can have some advantages from the point of view of the
photon detection and generation of novel quantum states, as was shown
recently in \cite{2011a,2011b}. But in that papers the problem was treated only
numerically, since the exact Rabi Hamiltonian does not allow for simple analytical solutions.
At the same time, it is desirable to have also at least approximate analytical solutions,
which could help us to understand better the mechanism and details of the
process. Finding such solutions and comparing them with numerical ones is the goal of this report.

\section{Analytical solutions for simplified Hamiltonians}

Following the studies \cite{Law94,DK96}, we choose the time dependence of
the cavity frequency in the harmonic form
$\omega (t)=\omega_0[1+\varepsilon \sin (\eta t)]$, where $\varepsilon \ll 1$
is the modulation amplitude and $\eta $ is the frequency of modulation.
It is known \cite{DK96} that in the absence of the atom-field interaction the
mean number of photons grows with time exponentially if $\eta \approx
2\omega_0$ and $\varepsilon\omega_0 t \gtrsim 1$.
%
Hereafter we normalize the unperturbed cavity frequency to $1$, writing the
modulation frequency as $\eta =2\left( 1+x\right) $, where $x$ is a small
resonance shift. Moreover, for $\varepsilon \ll 1$ we write $\omega \left(
t\right) \simeq 1$, as the modulation influence is only relevant for the
squeezing coefficient $\chi \left( t\right) \simeq \left( \varepsilon \eta
/4\right) \cos (\eta t)$.
Having in mind obtaining approximate analytical solutions, we replace
the Rabi Hamiltonian (\ref{HR}) by its Jaynes--Cummings reduced form \cite{JC}
$H_{JC}= g\left(a\sigma _{+} +a^{\dagger} \sigma _{-}\right)$.
Therefore, our starting point is the Hamiltonian
\begin{equation}
H_{0}=n+i\chi (t)(a^{\dagger 2}-a^{2})
+\frac{\Omega }{2}\sigma_{z}+
g(a \sigma _{+} +a^{\dagger }\sigma _{-})
\label{H0}
\end{equation}%
 in the weak atom-field coupling
regime, $|g|\ll 1,\Omega $.
We wish to find approximate analytical solutions to
the time-dependent Schr\"{o}dinger equation
$|\dot{\Psi}(t)\rangle=-iH_{0}|\Psi (t)\rangle $.
This problem was solved in \cite{PLA} under the restriction $|\varepsilon|\ll |g|$,
but now we consider the case of arbitrary (although small) values of $\varepsilon$ and $g$.

Note that we assume that $\chi(t)$ is the only time-dependent function,
while the coupling constant $g$ is time-independent.
In principle, photons can be created also in the case of fast variations of the coupling
constant $g$, instead of the cavity frequency $\omega$ \cite{Liber09}.
This case can be also interpreted as another realization of the DCE (in some wide sense).
Then the number of created photons can be, in principle, even bigger than in the
cavity DCE (since the time-dependent part of the Hamiltonian becomes linear
with respect to the annihilation/creation operators, instead of the quadratic Hamiltonian
(\ref{H00}), as was mentioned in \cite{revDCE}).
However, namely the cavity DCE seems the most impressive, from our point of view,
and for this reason we concentrate on it.
Formal mathematical relations between the cases $\{\omega(t),
g=const\}$ and $\{\omega=const, g(t)\}$ were discussed in \cite{2011a},
where the influence of dissipation was also taken into account.

The first step in obtaining analytical solutions is to go to the
interaction picture by means of the time-dependent unitary transformation
\[
|\Psi (t)\rangle =V(t)|\psi (t)\rangle , \quad
V\left( t\right) =\exp \left[ -it\left( \eta /2\right) \left( n+\sigma
_{z}/2\right) \right] .
\] 
Then the interaction Hamiltonian acting upon the new function $|\psi
(t)\rangle $ reads
\begin{eqnarray}
H &=&V^{\dagger }\left( t\right) H_{0}V\left( t\right) -~iV^{\dagger }\left(
t\right) \dot{V}\left( t\right)   \label{mus} \\
&=&g(a\sigma _{+}+a^{\dagger }\sigma _{-})+iq(a^{\dagger 2}-a^{2})-xn-\frac{%
\Delta +x}{2}\sigma _{z},  \notag
\end{eqnarray}%
where
$q\equiv \varepsilon \left( 1+x\right) /4$ and
$\Delta \equiv 1-\Omega $ is the detuning parameter.
In the following we consider the
separable initial state with the field in the vacuum state, and the atom in
an arbitrary pure state%
\begin{equation}
|\psi (0)\rangle =\left( \alpha |g\rangle +\beta |e\rangle \right) \otimes
|0\rangle ~,\qquad \beta =\sqrt{1-\alpha ^{2}},
\label{psiin}
\end{equation}%
and for simplicity we assume real $\alpha $ and $\beta $.
Explicit (although approximate) analytical expressions
for the probability amplitudes and the average values of the main system
observables can be obtained in two regimes: the dispersive and resonant ones.

\subsection{Dispersive regime}

The dispersive regime occurs when $\left\vert \Delta \right\vert \gg g$. In
this case we can simplify Hamiltonian (\ref{mus}) by means of the unitary
transformation $H^{\prime }\equiv UHU^{\dagger }$ with
\begin{equation}
U=e^{\zeta Y},\quad Y=a^{\dagger }\sigma _{-}-a\sigma _{+},\quad \zeta
=g/\Delta \ll 1.
\end{equation}%
Expanding $\exp \left( \zeta Y\right) $ in the Taylor series, we obtain%
\begin{eqnarray*}
H^{\prime } &=&i\theta \left( a^{\dagger 2}-a^{2}\right) -\varphi n-\varpi
\sigma _{z}-2iq\zeta \left( a^{\dagger }\sigma _{+}-a\sigma _{-}\right) \\
&&-\frac{\delta }{2}-\frac{4}{3}g\zeta ^{2}\left( an\sigma
_{+}+n\,a^{\dagger }\sigma _{-}\right) +g\mathcal{O}\left( \zeta ^{3}\right)
~,
\end{eqnarray*}%
where $\delta \equiv g^{2}/\Delta \ll 1$ is the dispersive shift and
\begin{equation*}
\varphi \equiv x+\delta \sigma _{z},~\theta \equiv q(1+\zeta ^{2}\sigma
_{z}),~\varpi \equiv (\Delta +\delta +x)/{2}~.
\end{equation*}%
The DCE is described by the term $i\theta \left( a^{\dagger 2}-a^{2}\right) $%
. In the absence of other terms, it would result in a slow evolution of the
photon operators (in the Heisenberg picture) on the time scale of the order
of $\varepsilon ^{-1}$. On the other hand, the term $\varpi \sigma _{z}$
alone would result in oscillations of the operators $\sigma _{\pm }$ with
the frequency $2\varpi $. Consequently, under the condition $|\varepsilon
|\ll |2\varpi |$ the terms containing operators $\sigma _{\pm }$ can be
considered as rapidly oscillating (and small due to the presence of
coefficients $\zeta $ or $\zeta ^{2}$), so that they can be removed
following the standard ideology of the rotating wave approximation. In this
way we arrive at the following effective Hamiltonian, which takes into
account the terms up to the second order in $\zeta $ (we neglect the
unessential constant term $-\delta /2$):
\begin{equation}
H_{ef}=i\theta \left( a^{\dagger 2}-a^{2}\right) -\varphi n-\varpi \sigma
_{z}~.
\end{equation}%
It is valid roughly for $|\delta |t\ll 1$. Then the time-dependent
wavefunction (in the interaction picture) can be written as
$
|\psi (t)\rangle =U^{\dagger }\hat{\Lambda}_{\varphi ,\theta }\exp (i\varpi
t\sigma _{z})U|\psi (0)\rangle $,
where we define the squeezing operator%
\begin{equation}
\hat{\Lambda}_{v_{1},v_{2}}\equiv \exp \left\{ \left[ iv_{1}n+v_{2}\left(
a^{\dagger 2}-a^{2}\right) \right] t\right\}.
\end{equation}%
Note that $\varphi ,\theta ,v_{1},v_{2}$ are operators with the respect to the
atomic basis, containing the diagonal matrix $\sigma _{z}$.
The operator $\hat{\Lambda}_{v_{1},v_{2}}$ has the following properties
\cite{Puri}:
\begin{equation*}
\hat{\Lambda}_{v_{1},v_{2}}^{\dagger }a\hat{\Lambda}_{v_{1},v_{2}}=\mathcal{C%
}_{v_{1},v_{2}}^{\ast }a+\mathcal{S}_{v_{1},v_{2}}a^{\dagger },
\end{equation*}%
\begin{equation*}
\Lambda _{v_{1},v_{2}}^{(n)}\equiv \langle 2n|\hat{\Lambda}%
_{v_{1},v_{2}}|0\rangle =\frac{e^{-iv_{1}t/2}}{\mathcal{C}%
_{v_{1},v_{2}}^{1/2}}\left( \frac{\mathcal{S}_{v_{1},v_{2}}}{\mathcal{C}%
_{v_{1},v_{2}}}\right) ^{n}\frac{\sqrt{\left( 2n\right) !}}{2^{n}n!},
\end{equation*}%
where $|k\rangle $ denotes the k-th Fock state and%
\begin{eqnarray*}
\mathcal{C}_{v_{1},v_{2}} &\equiv &\cosh \left( d_{v_{1},v_{2}}t\right)
-id_{v_{1},v_{2}}^{-1}v_{1}\sinh \left( d_{v_{1},v_{2}}t\right) , \\
\mathcal{S}_{v_{1},v_{2}} &\equiv &2d_{v_{1},v_{2}}^{-1}v_{2}\sinh \left(
d_{v_{1},v_{2}}t\right) ,~d_{v_{1},v_{2}}=\sqrt{4v_{2}^{2}-v_{1}^{2}}.
\end{eqnarray*}%
For $v_{1}=0$ we use the shorthand notation $\Lambda _{v_{2}}^{(n)}\equiv
\Lambda _{0,v_{2}}^{(n)}$, $\mathcal{C}_{v_{2}}\equiv \mathcal{C}%
_{0,v_{2}}=\cosh \left( 2v_{2}t\right) $, $\mathcal{S}_{v_{2}}\equiv
\mathcal{S}_{0,v_{2}}=\sinh \left( 2v_{2}t\right) $. After some
manipulations one can obtain the probability amplitudes (exact to the second
order in $\zeta $)%
\begin{eqnarray*}
\langle g,2n|\psi (t)\rangle &=&\alpha e^{-i\varpi t}\left( 1-\zeta
^{2}n\right) \Lambda _{\varphi _{-},\theta _{-}}^{\left( n\right) }, \\
\langle e,2n|\psi (t)\rangle &=&\beta \left\{ e^{i\varpi t}\left[ 1-\zeta
^{2}\left( n+1\right) \right] \Lambda _{\varphi _{+},\theta _{+}}^{\left(
n\right) }\right. \\
&&\left. +e^{-i\varpi t}\zeta ^{2}\left( 2n+1\right) \mathcal{C}_{\varphi
_{-},\theta _{-}}^{-1}\Lambda _{\varphi _{-},\theta _{-}}^{\left( n\right)
}\right\} , \\
\langle g,2n+1|\psi (t)\rangle &=&\beta \zeta \sqrt{2n+1}\left\{ e^{-i\varpi
t}\mathcal{C}_{\varphi _{-},\theta _{-}}^{-1}\Lambda _{\varphi _{-},\theta
_{-}}^{\left( n\right) }\right. \\
&&\left. -e^{i\varpi t}\Lambda _{\varphi _{+},\theta _{+}}^{\left( n\right)
}\right\} , \\
\langle e,2n-1|\psi (t)\rangle &=&\alpha e^{-i\varpi t}\zeta \sqrt{2n}%
\Lambda _{\varphi _{-},\theta _{-}}^{\left( n\right) },
\end{eqnarray*}%
where $\theta _{\pm }=q(1\pm \zeta ^{2})$ and $\varphi _{\pm }=x\pm \delta $.
Other relevant quantities as functions of time are as follows:
\begin{equation*}
\left\langle a\right\rangle =\zeta \alpha \beta \left[ \mathcal{S}_{\varphi
_{-},\theta _{-}}+\mathcal{C}_{\varphi _{-},\theta _{-}}^{\ast }-\Sigma _{0}\right],
\end{equation*}%
\begin{eqnarray*}
\left\langle a^{2}\right\rangle &=&\alpha ^{2}\left( 1-\zeta ^{2}\right)
\mathcal{S}_{\varphi _{-},\theta _{-}}\mathcal{C}_{\varphi _{-},\theta
_{-}}^{\ast }+\beta ^{2}\left\{ \mathcal{S}_{\varphi _{+},\theta _{+}}%
\mathcal{C}_{\varphi _{+},\theta _{+}}^{\ast }\right. \\
&&\left. +\zeta ^{2}\left(3\mathcal{S}_{\varphi _{-},\theta _{-}}\mathcal{C}%
_{\varphi _{-},\theta _{-}}^{\ast }-2\mathcal{S}_{\varphi _{-},\theta
_{-}}^{-1}\Sigma _{1}\right)\right\},
\end{eqnarray*}%
\begin{eqnarray*}
\left\langle n\right\rangle &=&\alpha ^{2}\left( 1-\zeta ^{2}\right)
\mathcal{S}_{\varphi _{-},\theta _{-}}^{2}+\beta ^{2}\left\{ \mathcal{S}%
_{\varphi _{+},\theta _{+}}^{2}
+3\zeta ^{2}\mathcal{S}_{\varphi _{-},\theta _{-}}^{2}\right. \\
&&\left. +2\zeta ^{2}\left(1-\mbox{Re}%
\left[\left(\Sigma _{0}+\Sigma _{1}\right)/\mathcal{C}_{\varphi _{-},\theta _{-}}^{\ast
}\right]\right)\right\},
\end{eqnarray*}%
\begin{eqnarray*}
P_{e} &\equiv &\left\vert \langle e|\psi (t)\rangle \right\vert ^{2}=\alpha
^{2}\zeta ^{2}\mathcal{S}_{\varphi _{-},\theta _{-}}^{2}
+\beta ^{2}\left\{1 -\zeta ^{2}\mathcal{S}_{\varphi _{+},\theta _{+}}^{2}\right.
 \\
&&\left.-2\zeta ^{2}\left(1-\mbox{Re}\left[\left(\Sigma _{0}+\Sigma _{1}\right)
/\mathcal{C}_{\varphi _{-},\theta_{-}}^{\ast }\right]\right)\right\},
\end{eqnarray*}%
where $\Sigma _{l}\equiv e^{2i\varpi t}\sum_{k=0}^{\infty }\left( 2k\right)
^{l}\Lambda _{\varphi _{-},\theta _{-}}^{(k)\ast }\Lambda _{\varphi
_{+},\theta _{+}}^{\left( k\right) }$ .

We see that the exponential photon generation from vacuum is possible only
if $d_{\varphi \pm ,\theta \pm }^{2}>0$ (accordingly with the atomic initial
state), therefore the resonance shift $x$ must be adjusted as function of
the atomic initial state. Thus, without the resonance adjustment, i.e.
setting $x=0$, there will be no exponential photon creation for $\theta
_{\pm }<\left\vert \delta \right\vert /2$, or roughly for $\varepsilon
<2\left\vert \delta \right\vert $. Moreover, for $x=\delta $ (or $x=-\delta
$), only the initial state $\alpha |g,0\rangle $ ($\beta |e,0\rangle $)
contributes to the exponential photon growth whenever $\varepsilon
<4\left\vert \delta \right\vert $.

In particular, for the initial state $|g,0\rangle $, by adjusting the
resonance shift to the most favorable condition $x=\delta $, one obtains the
following simple expressions:
\begin{equation}
P_{e}=\zeta ^{2}\left\langle n\right\rangle ,\quad \left\langle
n\right\rangle =\left( 1-\zeta ^{2}\right) \sinh ^{2}(2\theta _{-}t),
\label{bb1}
\end{equation}%
\begin{equation}
Q=1+2\left\langle n\right\rangle -\zeta ^{2},\quad g^{\left( 2\right)
}=3+\sinh ^{-2}(2\theta _{-}t),
\end{equation}%
\begin{eqnarray}
\left\langle (\Delta X)^{2}\right\rangle  &=&\frac{1}{2}\left[ \left(
1-\zeta ^{2}\right) e^{4\theta _{-}t}+\zeta ^{2}\right] , \\
\left\langle (\Delta P)^{2}\right\rangle  &=&\frac{1}{2}\left[ \left(
1-\zeta ^{2}\right) e^{-4\theta _{-}t}+\zeta ^{2}\right] ,  \label{bb4}
\end{eqnarray}%
where $Q=[\left\langle (\Delta n)^{2}\right\rangle -\left\langle
n\right\rangle ]/\left\langle n\right\rangle $ is the Mandel factor and $%
g^{\left( 2\right) }\equiv \left\langle a^{\dagger }a^{\dagger
}aa\right\rangle /\left\langle n\right\rangle ^{2}=1+Q/\left\langle
n\right\rangle $ is the second order coherence function. The field
quadratures (in the interaction picture) are defined as
\begin{equation*}
X=\left( a+a^{\dagger }\right) /\sqrt{2},\quad P=\left( a-a^{\dagger
}\right) /(\sqrt{2}i).
\end{equation*}

\subsection{Resonant regime}

In the resonant regime, $\Delta =0$, we choose the unitary operator $U$ in
the form%
\begin{equation}
U=e^{i\xi Y},\quad Y=a^{\dagger }\sigma _{+}+a\sigma _{-},\quad \xi
=2g/\varepsilon ~,
\end{equation}%
assuming the weak coupling regime, $\xi \ll 1$ (in the opposite case, $\xi
\gg 1$, no more than two photons can be created due to the fast exchange of
excitations between the field and atom \cite{PLA}). As shown in \cite%
{2011a,2011b}, in this case it is convenient to set $x=0$, so one obtains%
\begin{eqnarray*}
H^{\prime } &\simeq &\frac{\varepsilon }{4}\Big[i(1+\xi ^{2}\sigma
_{z})(a^{\dagger 2}-a^{2}) \\
&&-\frac{4}{3}\xi ^{3}\left( an\sigma _{+}-a^{\dagger 3}\sigma
_{+}+h.c.\right) +\mathcal{O}(\xi ^{4})\Big],
\end{eqnarray*}%
where $h.c.$ denotes the hermitian conjugate. Thus, to the second order in $%
\xi $ we obtain a slightly modified squeezing effective Hamiltonian%
\begin{equation}
H_{ef}=i\vartheta (a^{\dagger 2}-a^{2}),\quad \vartheta \equiv (\varepsilon
/4)(1+\xi ^{2}\sigma _{z}),
\end{equation}%
valid roughly for $gt\ll 1$. Then the wavefunction is%
\begin{equation*}
|\psi (t)\rangle =U^{\dagger }\hat{\Lambda}_{0,\vartheta }U|\psi (0)\rangle ,
\end{equation*}%
so that the probability amplitudes to the second order in $\xi $ read (here $%
\vartheta _{\pm }=(\varepsilon /4)(1\pm \xi ^{2})$ and $\vartheta _{0}\equiv
\varepsilon \xi ^{2}/2$):%
\begin{eqnarray*}
\left\langle g,2n-1|\psi (t)\right\rangle  &=&-i\beta \xi \sqrt{2n}\Lambda
_{\vartheta _{+}}^{\left( n\right) }, \\
\langle g,2n|\psi (t)\rangle  &=&\alpha \left\{ \left[ 1-\xi ^{2}\left(
n+1\right) \right] \Lambda _{\vartheta _{-}}^{\left( n\right) }\right.  \\
&&\left. +\xi ^{2}\left( 2n+1\right) \mathcal{C}_{\vartheta
_{+}}^{-1}\Lambda _{\vartheta _{+}}^{\left( n\right) }\right\} , \\
\left\langle e,2n|\psi (t)\right\rangle  &=&\beta (1-\xi ^{2}n)\Lambda
_{\vartheta _{+}}^{\left( n\right) }, \\
\langle e,2n+1|\psi (t)\rangle  &=&i\alpha \xi \sqrt{2n+1}\left( \mathcal{C}%
_{\vartheta _{+}}^{-1}\Lambda _{\vartheta _{+}}^{\left( n\right) }-\Lambda
_{\vartheta _{-}}^{\left( n\right) }\right) .
\end{eqnarray*}

The expressions for the average values of the main observables are as
follows:%
\begin{equation*}
\left\langle X\right\rangle =0,\quad \left\langle P\right\rangle =\sqrt{2}%
\xi \alpha \beta \left[ e^{-2\vartheta _{+}t}-\mathcal{C}_{\vartheta
_{0}}^{-1/2}\right] ,
\end{equation*}%
\begin{eqnarray*}
\left\langle a^{2}\right\rangle  &=&\frac{\alpha ^{2}}{2}\left\{ \mathcal{S}%
_{2\vartheta _{-}}+\xi ^{2}\left[ 3\mathcal{S}_{2\vartheta _{+}}-4\mathcal{S}%
_{\vartheta _{-}}\mathcal{C}_{\vartheta _{0}}^{-3/2}\right] \right\}  \\
&&+\frac{\beta ^{2}}{2}(1-\xi ^{2})\mathcal{S}_{2\vartheta _{+}}~,
\end{eqnarray*}%
\begin{eqnarray*}
\left\langle \sigma _{-}\right\rangle  &=&\alpha \beta \left\{ \mathcal{C}%
_{\vartheta _{0}}^{-1/2}+\xi ^{2}\left[ \mathcal{C}_{\vartheta
_{0}}^{-1/2}\left( \frac{1}{2}\mathcal{S}_{2\vartheta _{+}}-\mathcal{C}%
_{\vartheta _{-}}^{2}\right) \right. \right.  \\
&&\left. \left. -\mathcal{C}_{\vartheta _{0}}^{-3/2}\mathcal{S}_{\vartheta
_{0}}\left( \frac{1}{2}\mathcal{S}_{2\vartheta _{-}}+\mathcal{S}_{\vartheta
_{+}}^{2}\right) +e^{-2\vartheta _{+}t}\right] \right\} ,
\end{eqnarray*}%
\begin{eqnarray*}
\left\langle n\right\rangle  &=&\alpha ^{2}\left\{ \mathcal{S}_{\vartheta
_{-}}^{2}+2\xi ^{2}\left[ 1+\frac32\mathcal{S}_{\vartheta _{+}}^{2}-\mathcal{C}%
_{\vartheta _{-}}\mathcal{C}_{\vartheta _{0}}^{-3/2}\right] \right\}  \\
&&+\beta ^{2}(1-\xi ^{2})\mathcal{S}_{\vartheta _{+}}^{2},
\end{eqnarray*}%
\begin{equation*}
P_{e}=2\alpha ^{2}\xi ^{2}\left[ 1+\frac12\mathcal{S}_{\vartheta _{-}}^{2}-%
\mathcal{C}_{\vartheta _{-}}\mathcal{C}_{\vartheta _{0}}^{-3/2}\right]
+\beta ^{2}\left[1-\xi ^{2}\mathcal{S}_{\vartheta _{+}}^{2}\right].
\end{equation*}%
In this case simplified expressions can be obtained for the initial excited
state ($\alpha =0$): $P_{e}=1-\xi ^{2}\left\langle n\right\rangle $, while $%
\left\langle n\right\rangle $, $Q$, $g^{(2)}$, $\langle \left( \Delta
X\right) ^{2}\rangle $ and $\langle \left( \Delta P\right) ^{2}\rangle $ are
given by Eqs. (\ref{bb1})-(\ref{bb4}) with substitutions $\zeta \rightarrow
\xi $ and $\theta _{-}\rightarrow \vartheta _{+}$.

\section{Discussion and conclusions}

\begin{figure}[htb]
\begin{center}
\includegraphics[width=0.48\textwidth]{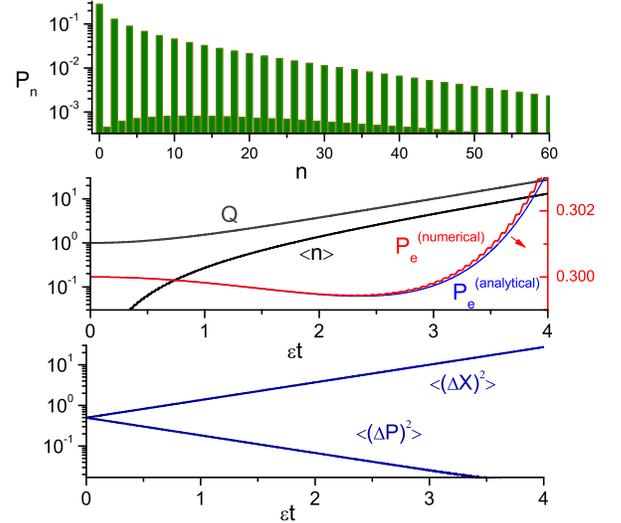} {}
\end{center}
\caption{Exact numerical results in the resonant
regime and the analytical results for $P_{e}$.
The photon number distribution $P_{n}\equiv |\langle n|%
\protect\psi \rangle |^{2}$ is evaluated for $%
\protect\varepsilon t=3.9$.}
\label{f1}
\end{figure}

Since the results of the two preceding sections were derived in the
frameworks of small-parameter expansions, we can believe that they are
correct on the time scales $t\ll 1/|\delta |$ in the dispersive regime and
for $t\ll 1/g$ in the resonance case (so that in both the cases the
 product $\varepsilon t$ can be bigger than unity, thus
enabling the generation of many photons). To check the validity of
analytical formulas obtained, we solved the Schr\"{o}dinger equation
numerically for the given initial conditions, using instead of
the {\em approximate\/} Hamiltonian (\ref{H0}) the exact
initial Hamiltonian $H_{00}+H_a +H_{R}$ [i.e., taking into the account
the interaction in the complete Rabi form (\ref{HR})]. The numerical results
turn out to be in a very good agreement with the analytical ones.
For example, in Fig. \ref{f1}
we show the exact dynamics for the initial state (\ref{psiin})
with $\beta ^{2}=0.3$ in the
resonant regime with parameters $\Delta =x=0$, $g=5\cdot 10^{-4}$, and $%
\varepsilon =2\cdot 10^{-2}$.
A small noticeable difference can be seen only for the evolution of
the probability $P_{e}(t)$ of finding the atom in the excited state
for $\varepsilon t >3$ (the variations of this quantity are small
due to the smallness of parameter $\xi=0.05$). The analytical results for other quantities are
indistinguishable  from the numerical ones within the thickness of lines
used in the plots. The mean number of photons grows exponentially for
$\varepsilon t >2$, and the photon statistics is super-Poissonian,
since the quantum state is close to the squeezed vacuum state
(small nonzero probabilites of the odd numbers of quanta arise just due to the
atom-field interaction).

\begin{figure}[htb]
\begin{center}
\includegraphics[width=0.48\textwidth]{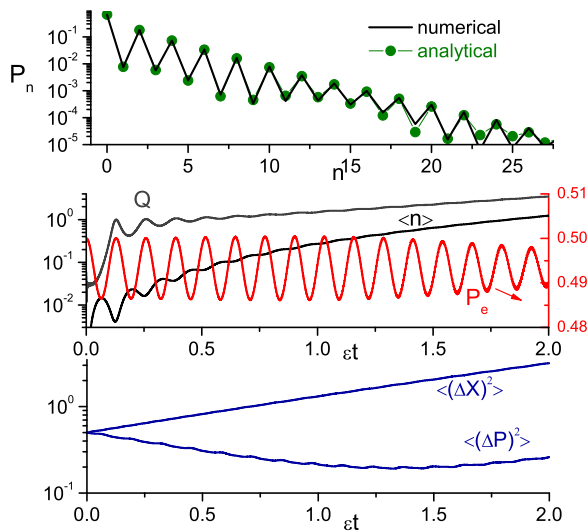} {}
\end{center}
\caption{Exact numerical results  in the dispersive
regime. $P_{n}$ is evaluated for $\protect\varepsilon t=2$,  both
analytically and numerically.}
\label{f2}
\end{figure}

In Fig. \ref{f2} we show analogous results for the initial
state (\ref{psiin}) with $\alpha ^{2}=1/2$ in the dispersive regime with parameters $x=0$, $%
g=5\cdot 10^{-3}$, $\Delta =12g$, and $\varepsilon =3\delta $, comparing
numerical and analytical values for the photon number distribution.
Again, the coincidence is very good, since the differences are seen only for
very low probabilities, less than $10^{-4}$. For other quantities,
analytical results are indistinguishable from numerical ones in all plots.

In conclusion, we obtained closed analytical expressions for the atom-field
dynamics generated by the dynamical Casimir effect in the resonant and
dispersive regimes, for arbitrary pure atomic initial state. Our results are
exact to second order in the small parameters $\zeta =g/\Delta $ and $\xi
=2g/\varepsilon $, being in good agreement with numerical data, so they can
be used to quantify the influence of the atom on the DCE for different
modulation frequencies and atom-cavity detunings. An interesting unsolved
problem is to extend the time interval of validity of results from the
scales of the order of $\varepsilon^{-1}$ or $g^{-1}$ up to the scales of
the order of $\varepsilon^{-2}$ or $g^{-2}$ (in the absence of field-atom
coupling some results were obtained in \cite{Sriv06,Mend11}).

\section*{Acknowledgment}
V.V.D.  acknowledges the partial support of CNPq (Brazilian
agency).

\end{document}